\documentclass[sigconf,authorversion,nonacm]{acmart}
\usepackage{xcolor}
\usepackage{tikz}

\usepackage{graphicx} 
\usepackage{subcaption}
\usepackage{fancybox}

\AtBeginDocument{%
  }
\usepackage[utf8]{inputenc}
\usepackage{float} 

\copyrightyear{2025} 
\acmYear{2025} 
\setcopyright{rightsretained}
\acmConference[CHI EA '25]{Extended Abstracts of the CHI Conference on Human Factors in Computing Systems}{April 26-May 1, 2025}{Yokohama, Japan}
\acmBooktitle{Extended Abstracts of the CHI Conference on Human Factors in Computing Systems (CHI EA '25), April 26-May 1, 2025, Yokohama, Japan}
\acmDOI{10.1145/3706599.3720078}
\acmISBN{979-8-4007-1395-8/2025/04}

\begin{document}

\title[Escaping the Filter Bubble]{Escaping the Filter Bubble: Evaluating Electroencephalographic Theta Band Synchronization as Indicator for Selective Exposure in Online News Reading}

\author{Thomas Krämer }
\email{thomas.kraemer@gesis.org}
\orcid{0000-0003-0507-7843}
\affiliation{
  \institution{GESIS Leibniz Institute for the Social Sciences}
  \city{Cologne}
  \country{Germany}
}

\author{Daniel Hienert }
\email{daniel.hienert@gesis.org}
\orcid{0000-0002-2388-4609}
\affiliation{
  \institution{GESIS Leibniz Institute for the Social Sciences}
  \city{Cologne}
  \country{Germany}
}

\author{Francesco Chiossi}
 \orcid{0000-0003-2987-7634}
\affiliation{
  \institution{LMU Munich}
  \city{Munich}
  \country{Germany}
}
\email{francesco.chiossi@lmu.de}

\author{Thomas Kosch}
\orcid{0000-0001-6300-9035}
\affiliation{
  \institution{HU Berlin}
  \city{Berlin}
  \country{Germany}}
\email{thomas.kosch@hu-berlin.de}

\author{Dagmar Kern}
\email{dagmar.kern@gesis.org}
\orcid{0000-0003-1794-625X}
\affiliation{
  \institution{GESIS Leibniz Institute for the Social Sciences}
  \city{Cologne}
  \country{Germany}
}
\renewcommand{\shortauthors}{Krämer et al.}

\begin{abstract}
Selective exposure to online news occurs when users favor information that confirms their beliefs, creating filter bubbles and limiting diverse perspectives. Interactive systems can counter this by recommending different perspectives, but to achieve this, they need a real-time metric for selective exposure. We present an experiment where we evaluate Electroencephalography (EEG) and eye tracking as indicators for selective exposure by using eye tracking to recognize which textual parts participants read and using EEG to quantify the magnitude of selective exposure. Participants read online news while we collected EEG and eye movements with their agreement towards the news. We show that the agreement with news correlates positively with the theta band power in the parietal area. Our results indicate that future interactive systems can sense selective exposure using EEG and eye tracking to propose a more balanced information diet. This work presents an integrated experimental setup that identifies selective exposure using gaze and EEG-based metrics. 
\end{abstract}



\maketitle

\section{Introduction and Background}

Selective exposure, the tendency of individuals to seek information that aligns with their pre-existing attitudes while avoiding contradictory information, is a major challenge to maintaining a diverse and informed public discourse \cite{Frey1986, boonprakong2025how}. This phenomenon contributes to the creation of echo chambers, where exposure to alternative viewpoints is limited, thereby reinforcing ideological divides between individuals. Quantifying the degree of selective exposure helps researchers to understand how individuals process viewpoints. In the context of Human-Computer Interaction (HCI), assessing selective exposure can help bias-aware systems to propose different viewpoints that differ from ideological stances~\cite{Dingler2023}. Traditional methods, such as self-reports, fail to fully capture the complexity of selective exposure. These approaches are limited in their ability to account for nuanced behaviors and contextual factors. Self-reports are prone to biases such as social desirability and confirmation bias, while log files lack the granularity needed to infer underlying cognitive processes~\cite{Zillich2019}. Moreover, experimental designs often rely on limited and artificially controlled stimuli, a ``captive audience problem'' \cite{Druckman2012}, which does not replicate the richness of real-world information environments~\cite{Yu2023}.

\begin{figure*}
    \centering
    \begin{subfigure}[b]{0.47\textwidth}
        \centering
        \includegraphics[width=\textwidth]{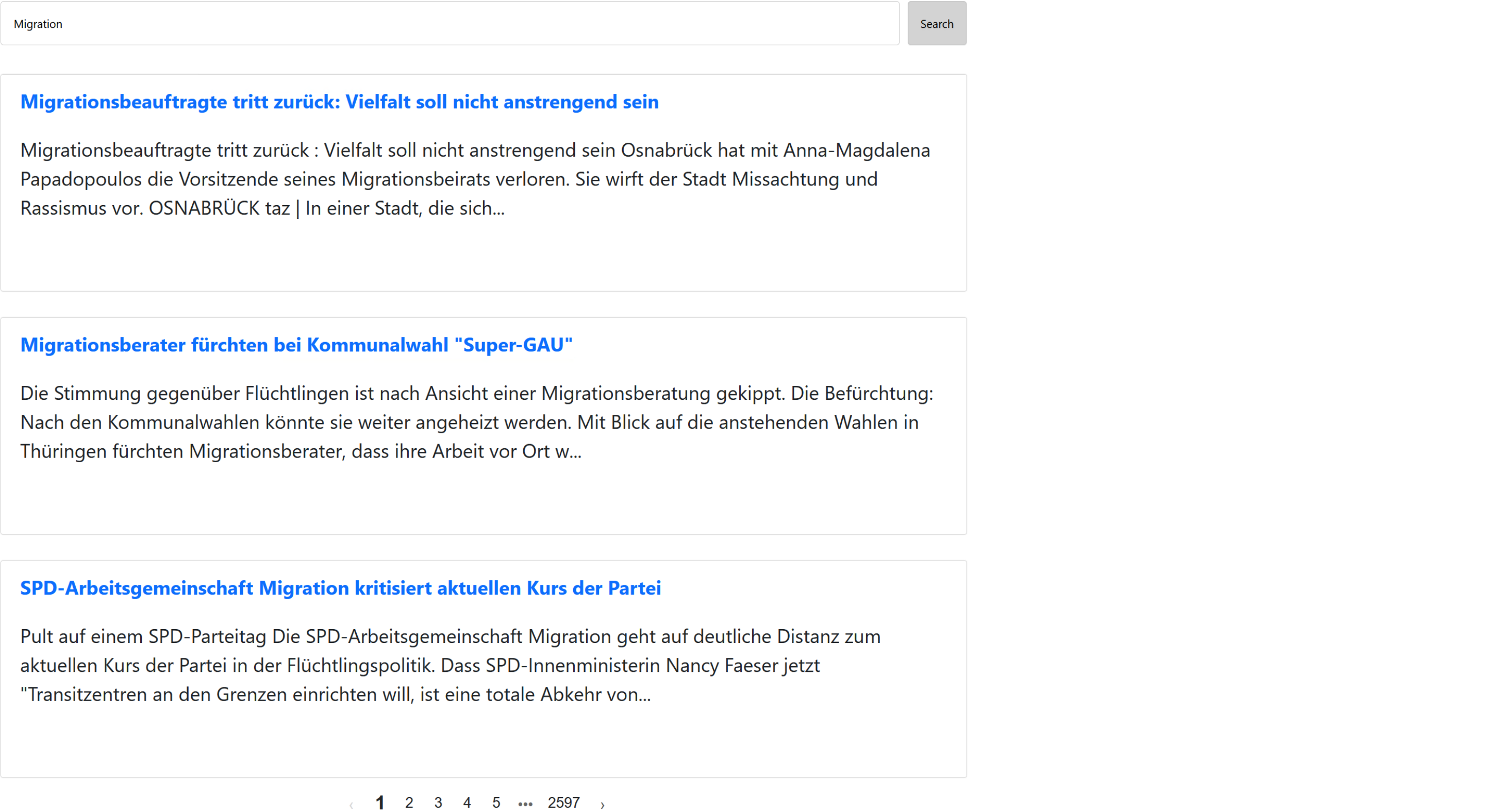}
        \caption{}
        \label{fig:news_search_system}
        \Description{}
    \end{subfigure}
    \hfill
    \begin{subfigure}[b]{0.47\textwidth}
        \centering
        \includegraphics[width=\textwidth]{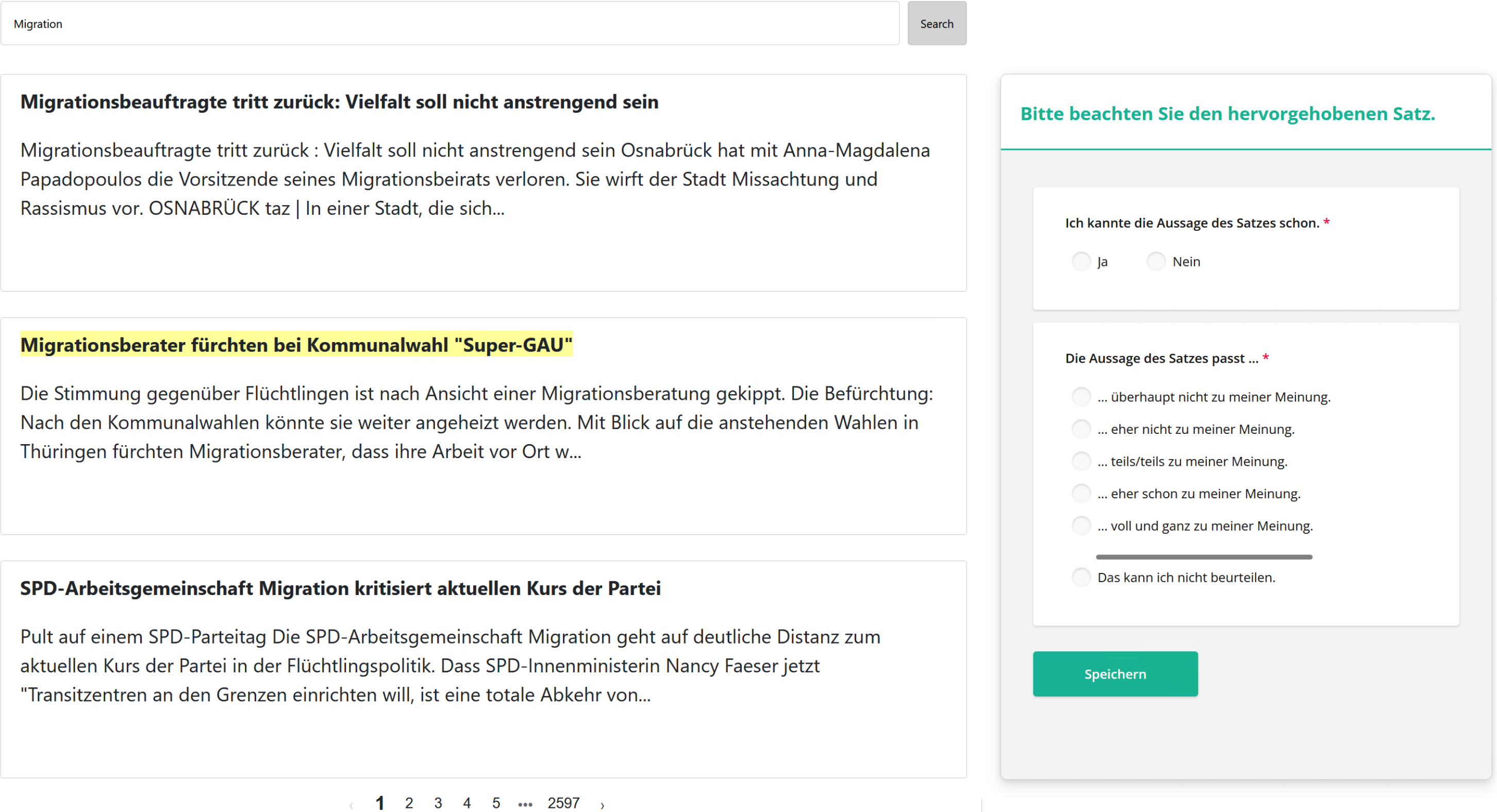}
        \caption{}
        \label{fig:news_search_system_with_feedback}
        \Description{}
    \end{subfigure}
    \caption{\textbf{(a):} News search system showing a result list containing German news articles for the query 'migration'. \textbf{(b):} The same view is used for user rating in the feedback phase with a highlighted sentence and a feedback form aside for sentence familiarity and agreement.}
    \Description{Left: News search system showing a result list for the query ``migration''. Right: The same view is used for data annotation in the feedback phase with a highlighted sentence and a feedback form aside for sentence familiarity and agreement.}
    \label{fig:news_news}
\end{figure*}

Previous research addressed selective exposure using behavioral metrics such as gaze duration and fixation patterns. For instance, Schmuck et al. \cite{schmuck2020avoiding} investigated selective exposure and avoidance using gaze data in response to attitude-consistent and attitude-discrepant stimuli. While these approaches provide valuable insights, gaze data alone cannot fully capture cognitive engagement, as fixations on a stimulus do not always imply attention or information processing \cite{foulsham2013mind}. Psychophysiological studies have explored the role of theta-band activity as a marker for memory encoding and decision-consistent processing \cite{Fischer2013, Klimesch2001, chiossi2025designing, long2024multimodal}, offering cortical physiological correlates for selective exposure. However, these studies focused on predefined static images or decision scenarios rather than reading self-selected information, which is a common scenario when consuming news or social media content~\cite{boonprakong2025assessing}.

To overcome these limitations, we developed a multimodal experimental framework combining Electroencephalography (EEG) and eye-tracking data to study selective exposure in a realistic web-browsing reading environment. Our study goes beyond existing work by analyzing both behavioral (gaze)~\cite{schneegassEtAl2020BrainCoDe} and physiological (EEG)~\cite{KoschEtAl2020OneDoesNotSimplyRSVP} signals to investigate how individuals process attitude-congruent and attitude-incongruent information while reading online news. Unlike previous unimodal research that relied on static visual stimuli~\cite{Fischer2013, Klimesch2001, schmuck2020avoiding}, our approach utilizes real news articles as text stimuli in a visually neutral presentation that can be searched and read naturally, making it closer to real-world information consumption. We propose the use of theta-band synchronization as a potential metric for selective exposure~\cite{Fischer2013}. Our results show EEG theta synchronization when participants read attitude-congruent sentences compared to attitude-discrepant ones. This work lays the groundwork for future studies aimed at developing near real-time, multimodal metrics for understanding selective exposure using real-time metrics.

\section{Experimental News Search System for Multimodal Assessment of Selective Exposure}

To investigate selective exposure in online news searches, we designed a system that collects and synchronizes eye tracking as well as EEG data. The primary goals of our system are twofold: (1) to understand user behavior during news reading and (2) to assess the feasibility of combining eye tracking and EEG for detecting selective exposure. To support these objectives, we implemented a custom news search system alongside a technical setup for data collection and synchronization. This section introduces the news search system and the associated technical setup.

\subsection{News Search System}
The core of our experimental design is a web-based news search application that allows participants to search, view, select, and read online news articles. Beyond enabling interaction with news content, the system also supports the administration of pre- and post-task questionnaires and the collection of user ratings at the sentence level. Participants interact with the system by entering keywords into a search bar, which generates a paginated result list displaying three items per page (see \autoref{fig:news_search_system}). Each result item includes the article title and the first sentence as a snippet. Clicking on an item opens the full article in a detailed view, from which participants can return to the result list via a back button. The paginated design minimizes scrolling, ensuring controlled interactions. To eliminate potential confounding variables, the application maintains a consistent and controlled presentation of text-based news content used in previous work~\cite{kosch2019investigating,KoschEtAl2020OneDoesNotSimplyRSVP}. Font size (24\,px), colors, and layout are uniform across the result list and detailed views, displayed on a 24-inch screen with a resolution of 1920\,$\times$\,1080 pixels. No images or additional styling elements are included, as graphical stimuli can manipulate cognitive processing compared to text stimuli. This controlled environment reduces external distractions that lead to noise in eye-tracking data. The news corpus comprises approximately 86,000 articles sourced from German online news outlets, including \textit{sueddeutsche.de}, \textit{zeit.de}, and \textit{bild.de}. These outlets were selected to represent diverse political orientations. Articles were collected using the NewsCatcherApi\footnote{\href{https://www.newscatcherapi.com}{https://www.newscatcherapi.com}} over a defined period (2023-12-19 to 2024-08-15), cleaned of artifacts, and indexed using Elasticsearch\footnote{\href{https://www.elastic.co/de}{https://www.elastic.co/de}}. By employing a large and diverse corpus, we address the ``captive audience problem'' \cite{Druckman2012} and simulate everyday news consumption behaviors.
\begin{figure*}
    \centering
    \includegraphics[width=1.0\linewidth]{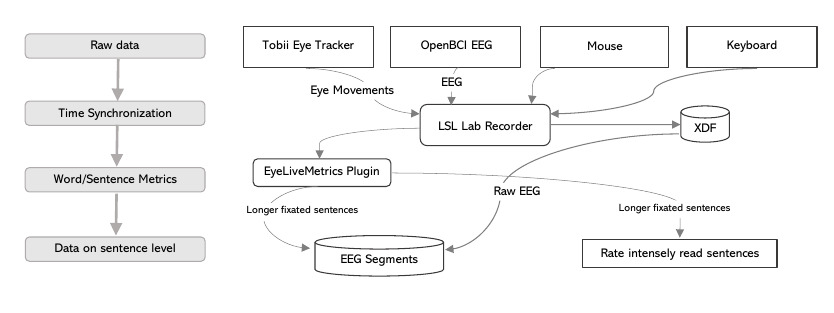}
    \caption{Process to collect different input streams to time-synchronized data and to compute sentence fixation times for the segmentation of EEG data and data annotation.}
    \label{fig:pipeline_technical_setup}
    \Description{TobiiEyeTracker, OpenBCI, Mouse, and Keyboard input streams are broadcast as raw data into LSL. The LSL Lab Recorder application stores the data in a single XDF file per participant and session, and also forward the gaze data to the EyeLiveMetrics browser plugin using WebSocket. EyeLiveMetrics backend extracts saccade and fixation metrics from the raw gaze data. It calculates the longest-fixated sentences, which are presented again to the participants to collect feedback. Further, the word fixation start and end timestamps are used to segment the EEG data from the XDF file.}
\end{figure*}
The system incorporates pre- and post-task questionnaires to collect sociodemographic data, personality traits, and attitudes toward the search task topic. After completing the search task, participants are prompted to provide feedback on specific sentences from the articles they read. Sentences for feedback can be selected based on predefined criteria, such as longer fixation durations (from gaze data) or significant deviations in EEG metrics relative to a baseline. Highlighted sentences are presented in their original context within the interface, and participants rate their familiarity with and agreement with the statement expressed in the sentence (see \autoref{fig:news_search_system_with_feedback}).

\subsection{Technical Setup}

The technical setup integrates input streams from eye tracking, EEG, mouse, and keyboard into a time-synchronized data output format. \autoref{fig:pipeline_technical_setup} illustrates the process. Fixation data from the eye tracker is used to segment EEG recordings, enabling analysis of neural activity corresponding to specific sentences. This multimodal approach provides a direct link between behavioral (gaze), cortical (EEG), and subjective (feedback) data, facilitating a comprehensive analysis of selective exposure mechanisms.

\label{sec:technical_setup}
A Tobii Spectrum eye tracker\footnote{\url{https://www.tobii.com/products/eye-trackers/screen-based/tobii-pro-spectrum}} is used for gaze tracking, at a sampling rate of 300 Hz. Electroencephalography is recorded with semi-dry electrodes with NaCl solution and an OpenBCI Cython Biosensing board with Daisy extension\footnote{\url{https://shop.openbci.com/products/all-in-one-gelfree-electrode-cap-bundle}}, with 16 EEG channels at 125 Hz, and additional reference and ground electrodes\footnote{\href{https://shop.openbci.com/cdn/shop/files/gelfree_electrode_cap_diagram_1020.png}{ https://shop.openbci.com/cdn/shop/files/gelfree\_electrode\_cap\_diagram\_1020.png for detailed positions}}. Impedance is kept below 20 k$\Omega$. EEG electrodes are mounted according to the 10-20-system \cite{Jasper1958}, and signal is recorded from Fp1, Fp2, F3, F4, Cz, Pz, P3, P4, T3, C3, C4, T4, T5, O1, O2 and T6 electrodes. Mouse and keyboard input is recorded at 1000 Hz. Signals of these four modalities are streamed into LabStreamingLayer \cite{Kothe2024}. Gaze data from the Tobii Spectrum eye tracker is streamed using the Tobii Pro Connector App\footnote{\url{https://github.com/labstreaminglayer/App-TobiiPro}}. OpenBCI GUI\footnote{\url{https://github.com/OpenBCI/OpenBCI_GUI}} is used for impedance checks and streaming EEG data into LSL. The keyboard/mouse connector app\footnote{\url{https://github.com/labstreaminglayer/App-Input}} is used for streaming keyboard strokes, mouse clicks, and mouse position into LSL. The LabRecorder app\footnote{\url{https://github.com/labstreaminglayer/App-LabRecorder}} combines all signal streams into a single data file in XDF format. Data of all modalities is aligned on a synchronized timeline\footnote{\href{https://labstreaminglayer.readthedocs.io/info/time_synchronization.html }{https://labstreaminglayer.readthedocs.io/info/time\_synchronization.html }} after import with the python pyxdf package\footnote{\url{https://github.com/xdf-modules/pyxdf}}.

In the next step, we use eye-tracking data to compute fixation times on single words and sentences. The EyeLiveMetrics browser plugin \cite{Hienert2024} captures raw eye-tracking coordinates and maps them to words as Areas of Interest (AOIs)\footnote{\url{https://git.gesis.org/iir/eyelivemetrics}}. For that purpose, we send a fork of the eye-tracking data LSL stream via a web socket into the EyeLiveMetrics browser plugin. Within EyeLiveMetrics, fixation, and saccades are classified and mapped to word AOIs. For each word of the whole web page, metrics like fixation start and fixation duration are computed and saved in a database. We accumulate word fixation data with an additional script and compute gaze-based metrics for sentences, e.g., fixation duration or fixation start and end times. This provides a direct relationship between the reading of specific text measured with eye tracking, the cognitive processing in the brain measured with EEG, and the rating of the sentence by the participant.

\section{Exploratory User Study}

We investigated the relationship between selective exposure and EEG theta synchronization in an exploratory user study. Participants were exposed to news articles and, later on, provided their agreement level to selected sentences. With this information, we were able to analyze the neural correlates of agreement systematically.

We selected immigration policy as the study topic due to its societal relevance and ability to elicit strong attitudes. Immigration policy was ranked as the most important issue in German pre-election polls in early 2025\footnote{\url{https://www.infratest-dimap.de/umfragen-analysen/bundesweit/ard-deutschlandtrend/2025/januar}}. This provoked high engagement and polarized responses from participants, facilitating the study of selective exposure. Our institution’s ethics committee approved the study.
\begin{figure*}[t]
    \centering
    \includegraphics[width=\textwidth]{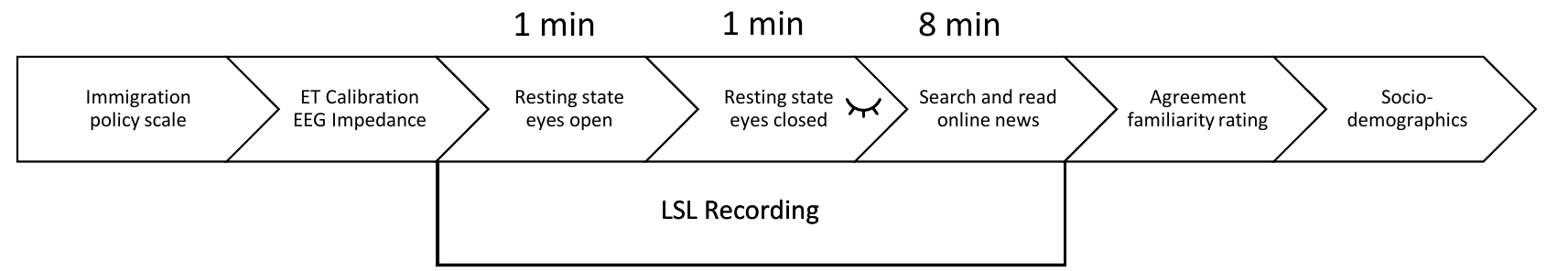}
    \caption{Experimental procedure of the study. Participants provided informed consent, completed an immigration policy scale \cite{Mau2023}, and answered questions on familiarity and interest in immigration policy. EEG and eye-tracker calibration followed. After resting-state EEG recording, participants performed an 8-minute news search on immigration policy. Post-task, they rated familiarity and agreement with their longest-fixated sentences, followed by socio-demographic questions.}
    \label{fig:procedure}
    \Description{A procedural flowchart illustrating the experimental design for studying selective exposure using EEG and eye tracking (ET) methods, showing sequential phases including ET calibration and EEG impedance (1 min), resting state with eyes open (1 min), resting state with eyes closed (1 min), searching and reading online news (8 min) with LSL recording, agreement and familiarity rating, and socio-demographics collection.}
\end{figure*}
\subsection{Procedure}
After briefing the participants on the study, they agree to participate by signing informed consent.
The experimenter then explains the task and the news corpus, mounts the EEG, checks electrode impedance, and starts the calibration of the eye tracker. 
After a successful calibration, a resting state EEG is recorded (one minute with eyes open while looking at a fixation cross and one minute with eyes closed). Then, the following task description is shown to the participants:

\addvspace{1em} 
   \textit{Your train is delayed, and you now have some unexpected free time. You decide to catch up on the current debate about immigration. Use the news search and enter any search terms you like. \newline Start your search on the topic of immigration now.}

\addvspace{1em} 

We intentionally kept the task description generic and did not induce a specific objective other than 'inform oneself' on immigration policy using the news search. No time limit is given in advance to avoid putting pressure on participants. However, the experimenter ends the search session after 8 minutes, removes the EEG, and stops the recording. Afterward, the participants are guided through rating the 19 longest-fixated sentences in their session. For that, they are shown the search result lists and the detailed views again, with the longest-fixated sentences highlighted in yellow, one after the other. For each highlighted sentence, they are asked to rate their familiarity with the statement ("I already knew the statement expressed in the sentence" yes/no) and their agreement towards that sentence on a 5-point Likert scale (from 1 = "The statement of the sentence fits not at all with my opinion." to 5 = "The statement of the sentence fits completely with my opinion."). Participants can also select the option "I cannot assess that.". The session closes with questions on their socio-demographics being filled out. Overall the experiment lasted one hour. Figure \ref{fig:procedure} illustrates the whole process.

\subsection{Multimodal Data Recording}

\subsubsection{Eye tracking Recording \& Preprocessing}
Gaze data were collected using the Tobii Spectrum eye tracker (300 Hz). Participants were seated approximately 60 cm from the eye tracker. Calibration was performed using a 6-point grid to ensure accurate gaze mapping.
To be able to present the longest-fixated sentences to the participants, the eye tracking data have to be processed (compare Section \ref{sec:technical_setup}). The EyeLiveMetrics plugin collects and registers fixation duration, on- and offset of words while participants are reading in the task. After the task, fixation duration sums with on- and offsets are computed for all sentences viewed. We define a sentence-level metric,  taking the summed fixation duration on all sentence words and dividing it by the count of words in the sentence. Only sentences with such normalized word fixation durations > 20 ms are considered. This filters out sentences with many skipped words. Sentences are sorted in descending order of sentence fixation duration sum. Output are longest-fixated sentences per participant. Based on that, we segment the EEG by using the fixation onsets of each word in the longest fixated sentences

\subsubsection{EEG Recording \& Preprocessing}
EEG data were recorded using semi-dry electrodes with a sodium chloride (NaCl) solution connected to an OpenBCI Cyton Biosensing Board as described in \ref{sec:technical_setup}. 
Continuous EEG data were segmented into epochs time-locked to fixation onset for each word in the longest-fixated sentences. Sentences were classified into two conditions: \textit{High Agreement}, which included sentences rated 4 or 5 on a 5-point Likert scale, and \textit{Low Agreement}, which included sentences rated 1 or 2. 

We preprocessed EEG data within MNE-Python \cite{gramfort2013meg}. First, a 50 Hz notch filter was applied to remove line noise, and a .5–15 Hz bandpass filter was used to eliminate high- and low-frequency noise. Signals were re-referenced to the common average reference. Independent Component Analysis was performed using the Infomax algorithm to identify and remove artifacts such as blinks, eye movements, and muscle activity. Artifact-prone components were automatically labeled and excluded using the ICLabel plugin integrated within MNE-Python \cite{pion2019iclabel, li2022mne}. Only epochs free of residual artifacts and corresponding to correct responses with a fixation on the target word were included in the final analysis, while error trials and those with distractor fixations were excluded. 

\begin{figure}
    \centering
    \includegraphics[width=1\columnwidth]{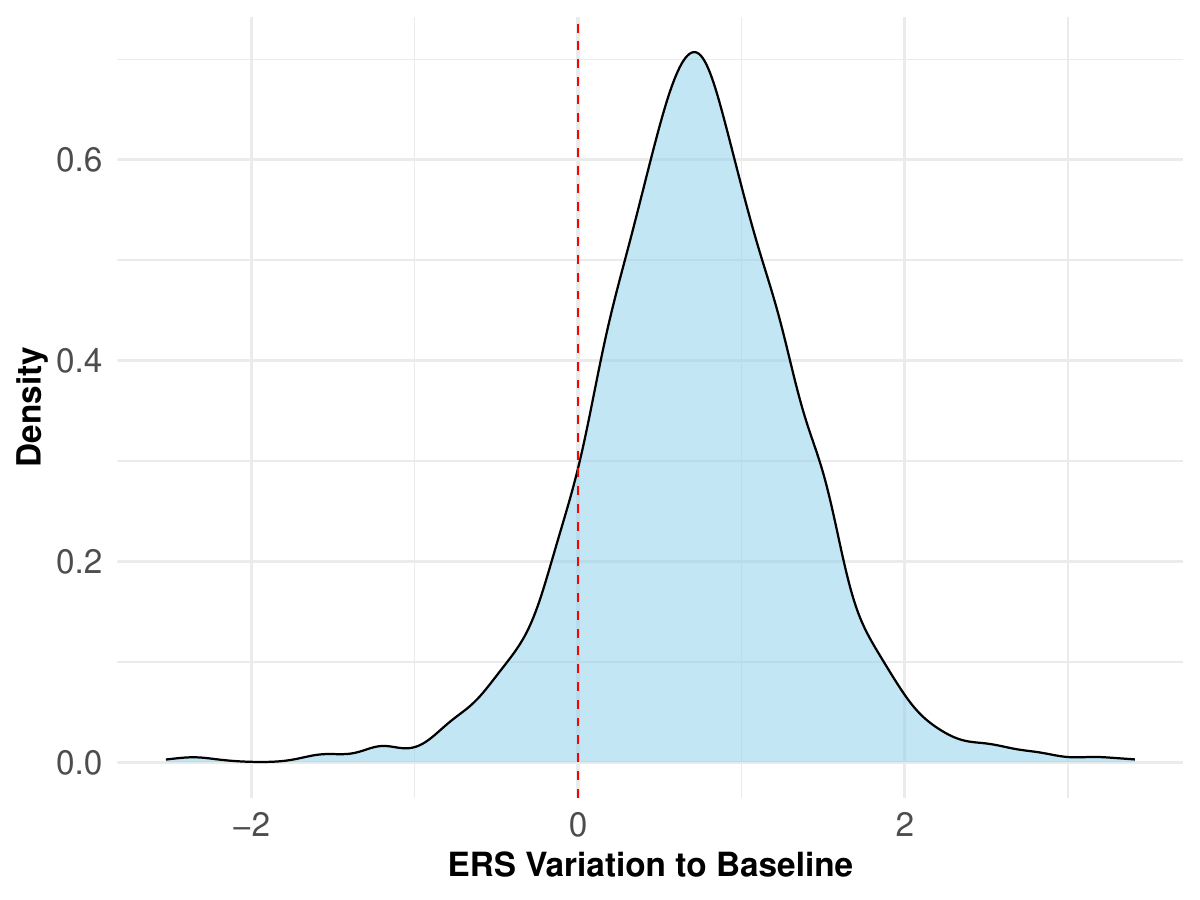} 
    \caption{Posterior distribution of the Event-Related Synchronization/Desynchronization (ERS/ERD) difference between high and low agreement conditions in the parietal region (1000–1500ms). The distribution peak at 0.71 SD power change with 89\% of the mass above zero (red dashed line) indicates stronger synchronization for high versus low agreement, with a Bayes Factor of 7.8 supporting this difference.}
    \label{fig:bayes_results_posterior}
    \Description{Posterior distribution of the difference in theta ERS between words in sentence with high and low agreement. The x-axis represents the  variation in ERS relative to baseline, and the y-axis displays density. A density curve shows posterior samples. A shift of the density away from zero suggests a credible difference in theta ERS between conditions.}
\end{figure}

EEG frequency analyses examine the spectral properties of EEG signal oscillations, including frequency, amplitude, and phase. Frequency is measured in Hertz (Hz) and indicates the number of full oscillations per second, while amplitude or power represents the voltage range between the maximum and minimum points. In this work, we investigated power changes in the theta band (5–8 Hz).
For fixated words, we computed theta power relative to a baseline period (-500 ms to 0 ms) preceding the first-word fixation onset as in previous work \cite{cona2020theta}. Theta power was extracted from electrodes P3, P4, and Pz based on previous work \cite{Klimesch2001, Klimesch2005, Fischer2013}.

We normalized data using z-scores, in order to account for between-subject and electrode variability, where \(\overline{\text{baseline}}_{\text{f}}\) represents the mean theta power during the baseline period (-500 ms to 0 ms), and \(\text{activity}_{\text{tf}}\) corresponds to theta power within the post-stimulus window (0 ms to 1500 ms). 

\[
Z_{\text{tf}} = \frac{\text{activity}_{\text{tf}} - \overline{\text{baseline}}_{\text{f}}}
{\sqrt{\frac{1}{N} \sum_{i=1}^{N} (\text{baseline}_{\text{f},i} - \overline{\text{baseline}}_{\text{f}})^2}}
\]

\subsection{Participants}
We recruited 8 participants. All participants have either normal or corrected vision (max. +/- 4.5 diopters) and are German native speakers. Their participation was compensated by 40 euros. The data of 3 participants had to be excluded because of unusable EEG data (1) or missing gaze data (2). 
Therefore, in the following, we only report on the results of the 5 remaining participants (2 male, 3 female, mean age 24 years).

\subsection{Results}
Regarding agreement with a sentence, we gained ratings for 65 sentences, whereby 33 are considered as high agreement (judged by 4 and 5 on the 5-point Likert scale) and 19 as low agreement (judged by 1 and 2 on the 5-point Likert scale).

\begin{figure}
    \centering
    \includegraphics[width=1\columnwidth]{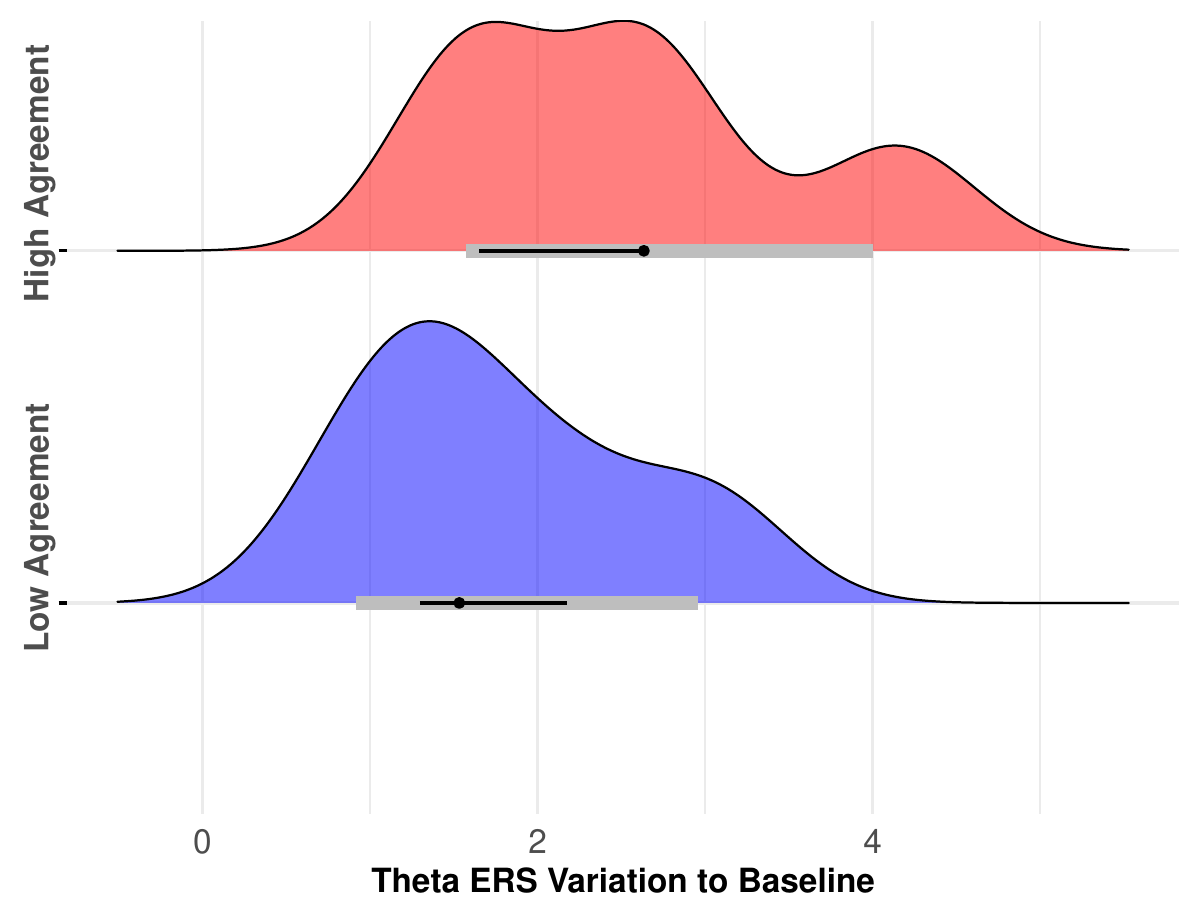} 
    \caption{Density ridgeline plot showing the distribution of mean Theta ERS/ERD in the parietal region for Low Agreement and High Agreement conditions, calculated over the 1000–1500ms time window. Median values and 80\%/95\% credible intervals are displayed.}
    \label{fig:bayes_results_conditions}
    \Description{The figure shows a ridge plot displaying the posterior distribution of theta Event-Related Synchronization (ERS) variation to baseline under low agreement and high agreement conditions. The x-axis represents the theta ERS variation to baseline, ranging from 0 to 6 standard deviations. The y-axis labels the low agreement (blue) and high agreement (red) conditions. The median values as dots and 80\%/95\% credible intervals as lines are displayed in black and gray colors, for both ridgeline plots. The vertical line at x = 0 serves as a reference for interpreting the credible intervals. The credible intervals do not fully overlap, medians are at different locations, suggesting that the high agreement condition is credibly associated with positive theta ERS changes relative to baseline.}
\end{figure}

\begin{figure*}
    \centering
    \includegraphics[width=0.7\textwidth]{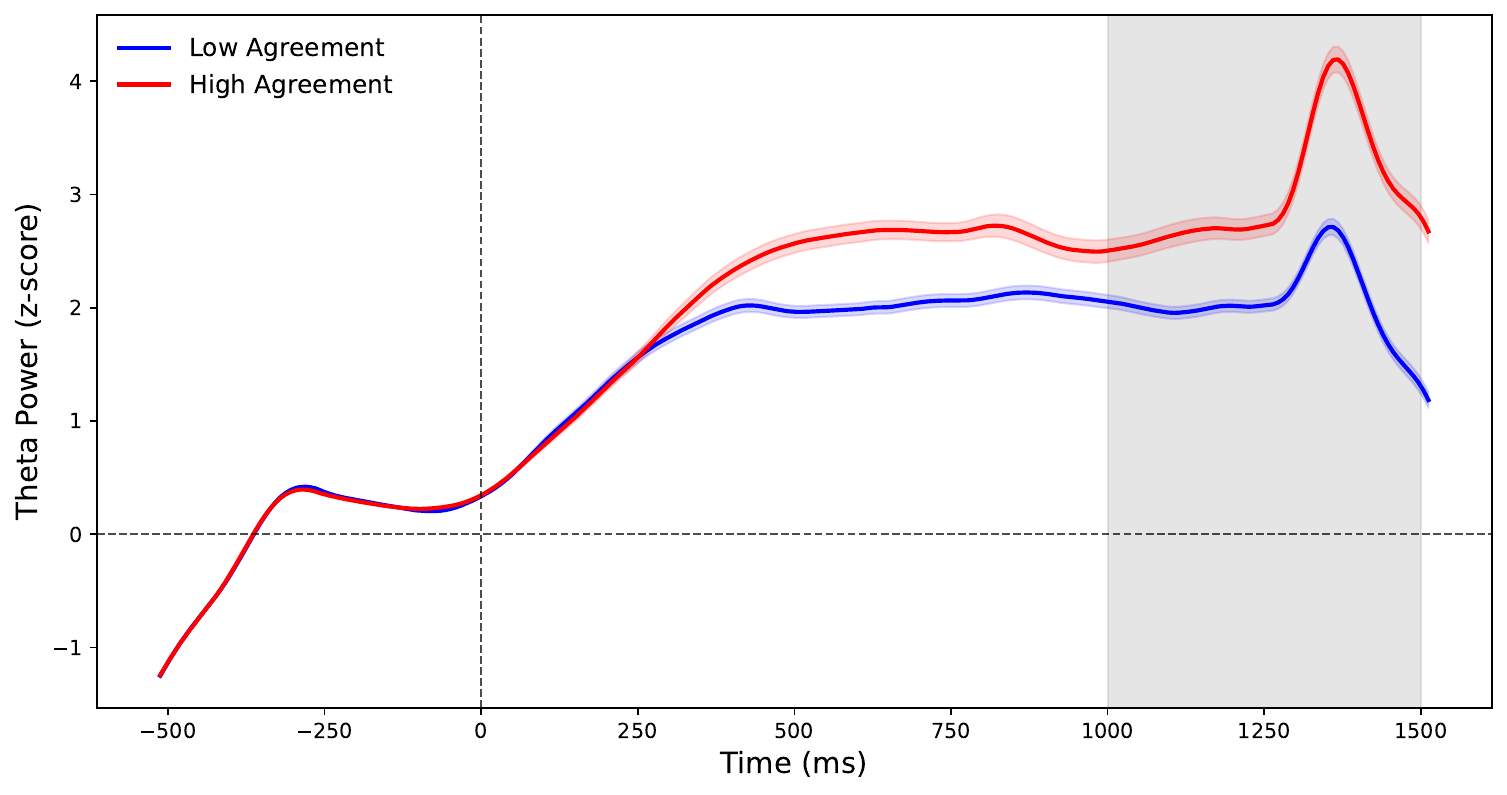}
    \caption{Event-related theta response for low and high agreement conditions. The shaded region (1000–1500 ms) marks the analysis window, where we found moderate evidence of greater theta synchronization with high agreement.}
    \label{fig:results_ers_theta}
    \Description{Changes of z-score normalized theta power for a word in sentence users reported to strongly agree or disagree with. Starting at around 250ms the change is stronger for sentences with high agreement.}
\end{figure*}

A Bayesian t-test on average z-score normalized theta power from the parietal region (P3, P4, Pz) during the 1000 - 1500ms time window revealed moderate evidence for a difference between high and low agreement conditions ($M = 0.72$, 95\% CI [$-0.32$, $1.73$], evidence ratio = 7.8, posterior probability = .89). This indicates that the data are approximately 7.8 times more likely under the alternative hypothesis than under the null hypothesis. \autoref{fig:bayes_results_posterior} and  \autoref{fig:bayes_results_conditions} show the distribution of the results of the event-related theta synchronization.  \autoref{fig:results_ers_theta} shows the event-related theta synchronization for low and high agreement.

\section{Discussion and Future Work}
In this work, we evaluated EEG and eye tracking as a research method for quantifying selective exposure in real-time. We discuss the implications of our research in the following.

Interactive systems that mitigate selective exposure can leverage our multimodal data approach, such as EEG and eye-tracking metrics, to foster a balanced and diverse information consumption. Yet, one limitation of our work is that we propose a research method using eye tracking and EEG, two metrics that are not common within consumer devices. However, researchers can use our method to study the degree of selective exposure within interactive systems in near real-time. Furthermore, the multimodal integration of EEG and eye tracking is becoming more dominant in virtual reality devices, which commonly contain eye tracking sensors and can be extended by EEG electrodes and other peripheral physiological measures~\cite{9756768, 10.1145}. 
In future work, we will evaluate our approach in immersive environments with consumer sensors in integrated virtual reality systems. Overall, we envision recommendation algorithms to adapt and reduce the reinforcement of echo chambers and encourage exposure to attitude-inconsistent information, ultimately reducing the spread of one-sided information and reducing susceptibility to misinformation and disinformation. For example, a chatbot can cognitively augment the user's sense of disinformation by sensing selective exposure in real-time and providing alternative views when reading news. Cognitive augmentation has been a recurring theme in HCI research and should be studied extensively when being used to reduce exposure to disinformation~\cite{VILLA2023107787, 10.1145/3529225}. To this end, our work contributes to the realization of bias-aware systems as proposed by Boonprakong et al.~\cite{Dingler2023} in the context of HCI.

The development of systems that modulate selective exposure raises significant ethical concerns.
Imposing diverse viewpoints without user consent might lead to perceived coercion, undermining trust in bias-aware systems~\cite{lee2004trust}. Designers of these systems must ensure transparency by clearly communicating how and why content is adapted. Additionally, for future work it is important to address the potential biases introduced by algorithmic interventions, as they may disproportionately affect users with specific cognitive or ideological profiles. 

\section{Conclusion}
\label{conclusion-future-work}
Our exploratory study suggests theta band power changes during news reading might contribute to an operationalization to detect selective exposure. We combined EEG and eye tracking with natural reading behavior to address methodological limitations of self-reporting and artificial reading tasks. In a preliminary study with five participants reading news from various political spectrums, we found moderate evidence for theta power as a predictor for high agreement with consumed information. Consequently, theta power is a promising metric for assessing selective exposure in real-time for interactive systems. Although our systematic approach requires laboratory hardware in experimental environments, we are confident that consumer devices will integrate EEG and eye tracking in future devices (e.g., virtual reality headsets). This lays the groundwork for future research to examine individual differences in theta responses and explore additional neurophysiological and gaze markers to develop robust metrics for detecting selective exposure during real-world information consumption.

\begin{acks}
 
This work is funded by the German Research Foundation (DFG) project ``Overcome Selective Exposure in Web Search by Considering Eye Movements and Physiological Signals'' (Project-ID~525041402). Furthermore, this work is supported by the DFG CRC 1404: ``FONDA: Foundations of Workflows for Large-Scale Scientific Data Analysis'' (Project-ID 414984028) and DFG CRC. This research is also supported by DFG -- Project-ID~251654672 -- TRR~161.

\end{acks}

\bibliographystyle{ACM-Reference-Format}
\bibliography{lbw}

\end{document}